\documentclass[doublecol]{epl2}

\title{Coherent Hydrodynamic Coupling for Stochastic Swimmers}
\shorttitle{Coherent Hydrodynamic Coupling for Stochastic Swimmers} 

\author{A. Najafi\inst{1} \and R. Golestanian\inst{2}}
\shortauthor{A. Najafi and R. Golestanian}

\institute{
  \inst{1} Department of Physics, Zanjan University, Zanjan 313, Iran\\
  \inst{2} Department of Physics and Astronomy, University of Sheffield, Sheffield S3 7RH, UK
}

\abstract{A recently developed theory of stochastic swimming is used to study the notion of
coherence in active systems that couple via hydrodynamic interactions. It is shown that
correlations between various modes of deformation in stochastic systems play the same role as
the relative internal phase in deterministic systems. An example is presented where a simple
swimmer can use these correlations to hunt a non-swimmer by forming a hydrodynamic bound state
of tunable velocity and equilibrium separation. These results highlight the significance of
coherence in the collective behavior of nano-scale stochastic swimmers.}

\begin{document}

\maketitle

\section{Introduction}
Swimming strategies for microorganisms and microbots need to take
into account the peculiarities that arise in low Reynolds number hydrodynamics
\cite{PoLa-rev,Taylor,Purcell}. When utilizing only a small number of degrees
of freedom, a careful non-reciprocal prescription of cyclic deformations
is needed to achieve swimming \cite{shapere,becker,3SS,josi,drey,feld,peko,yeomans1,Yeomans2,Karsten,3SS-perturb,zargar}.
While these ideas have been primarily developed to describe the swimming of bacteria \cite{BergEC},
sperms \cite{Kruse}, and other micro-scale living systems \cite{bray},
in recent years they have attracted additional interest with the advent of the first
generation of artificial microswimmer prototypes \cite{exp-swim}.
Swimmers of micro- or nanoscale need to face an additional challenge, namely, the
overwhelming fluctuations that would act against their targeted mechanical task.
In its most basic form, the effect of fluctuations on the motion of swimmers that are
not directionally constrained or steered is to randomize their orientation via rotational
diffusion \cite{Darnton,gla,jon,Vlad,Dunkel}. The fluctuations can also interfere with the
propulsion mechanism and alter the swimming velocity \cite{PeMo}, for example via the density
fluctuations in the case of self-phoretic swimmers \cite{msd} or fluctuations in the
conformational changes in deforming swimmers \cite{RG+AA}.

Interacting swimmers \cite{ped-rev} are known to have rather complex many-body behaviors
\cite{simulation}, which can be understood in terms of instabilities in the context of
continuum theories (that are constructed based on symmetry considerations) \cite{instability}.
Another fascinating consequence of long-range hydrodynamic interactions between active
objects with cyclic motions is the significance of {\em internal phase} as a key
dynamical variable \cite{Yeomans2,Yeomans3,denis}, and the possibility of synchronization
\cite{sync}. However, most current theoretical studies of the collective behavior of swimmers
(continuum theories and simulations) ignore the possible effects of coherence, and it is
natural to wonder if this is justified.

One could argue that the overwhelming fluctuations that are present at small scale may
wash out any trace of coherence among swimmers. To examine the validity of this argument,
we consider the following question: does the notion of relative internal phase apply to
stochastic swimmers? We use a statistical description to model the dynamics of systems
that undergo random conformational changes while interacting hydrodynamically. Using a specific
example of a three-sphere system coupled to a two-sphere system, we calculate the swimming
velocities as functions of the statistical transition rates for the conformational changes.
We show that coherence could be introduced in the system through the correlations between
the deformations, and that it can be used to create a stable bound state between the three-sphere
swimmer and the two-sphere system (that cannot swim when isolated) with a tunable equilibrium
distance and velocity.

\section{Hydrodynamic Model}
\begin{figure}[t]
\includegraphics[width=.99\columnwidth]{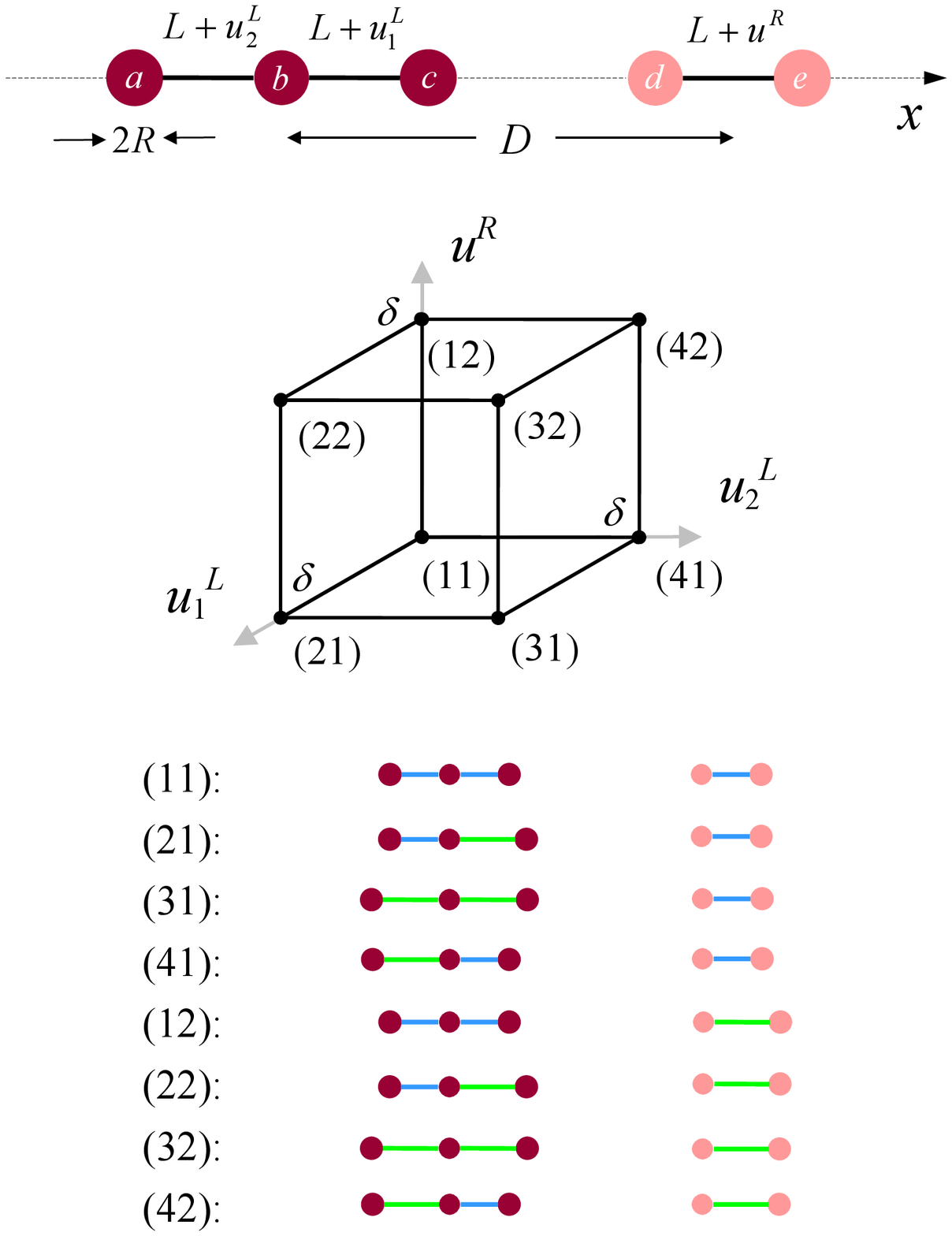}
\caption{Schematic view of a three-sphere swimmer interacting
with a two-sphere system (top) and the three dimensional configuration space (middle)
representing the eight possible distinct conformational states of the combined system (bottom).
We denote by $(i \alpha)$ the state where the three-bead system is in state $i$ and
the two-bead system is in state $\alpha$.}
\label{fig:schem1}
\end{figure}

Consider the a three-sphere system that is located
collinearly at a distance from a two-sphere system, as shown schematically in
Fig. \ref{fig:schem1}. The two systems undergo conformational changes by opening
and closing of the three arms, which could only lead to net swimming for the
three-sphere system (and not the two-sphere system) when isolated, due to
scallop theorem \cite{Purcell}. When at a finite distance $D$, the two systems
interact hydrodynamically, and their dynamics will be coupled to each other.
The system on the left (Fig. \ref{fig:schem1}) is made up of three spheres of
radius $R$ that are connected by arms of lengths $L+u_{1}^{L}(t)$ and $L+u_{2}^{L}(t)$,
while the system on the right consists of two similar spheres connected by an
arm of length $L+u^{R}(t)$. For simplicity, we assume that the linker do not
interact with the fluid. To Analyze the dynamics of the system, we use the
linearity of the Stokes equation---the equation for hydrodynamics in zero Reynolds
number---and express the velocity of each sphere $v_i$ as a linear
combination of the force $f_j$ acting on a different sphere $j$:
\begin{equation}
v_i=\sum_{j=a}^{e}M_{ij}f_j,\label{eq:oseen}
\end{equation}
where the details of the hydrodynamic interactions are entailed in the
coefficients $M_{ij}$. Using Oseen's approximation, we can write simple closed form
expressions for the coefficients when the spheres are considerably far from each other.
Denoting the positions of the spheres by $x_i$, we have:
\begin{equation}
M_{ij}=\left\{
\begin{array}{l}
\frac{1}{6\pi\eta R},~~ i=j, \nonumber\\
\\
\frac{1}{4 \pi \eta |x_{i}-x_{j}|},~~ i \neq j, \nonumber\\
\end{array}
\right.\label{eq:M-def}
\end{equation}
where $\eta$ is the viscosity of the fluidic medium.
Equation (\ref{eq:oseen}) thus gives us five equation for the ten unknowns $v_i$ and $f_i$
($i=a, \cdots, e$). Maintaining force-free conditions on the two systems, namely,
$f_a+f_b+f_c=0$ and $f_d+f_e=0$, provides two additional equations. The final three equations
are obtained by the kinematic constraints $v_b-v_a=\dot{u}_{2}^{L}$, $v_c-v_b=\dot{u}_{1}^{L}$, and $v_e-v_d=\dot{u}^{R}$, where the dot denotes $d/dt$.

\begin{figure}[t]
\includegraphics[width=.80\columnwidth]{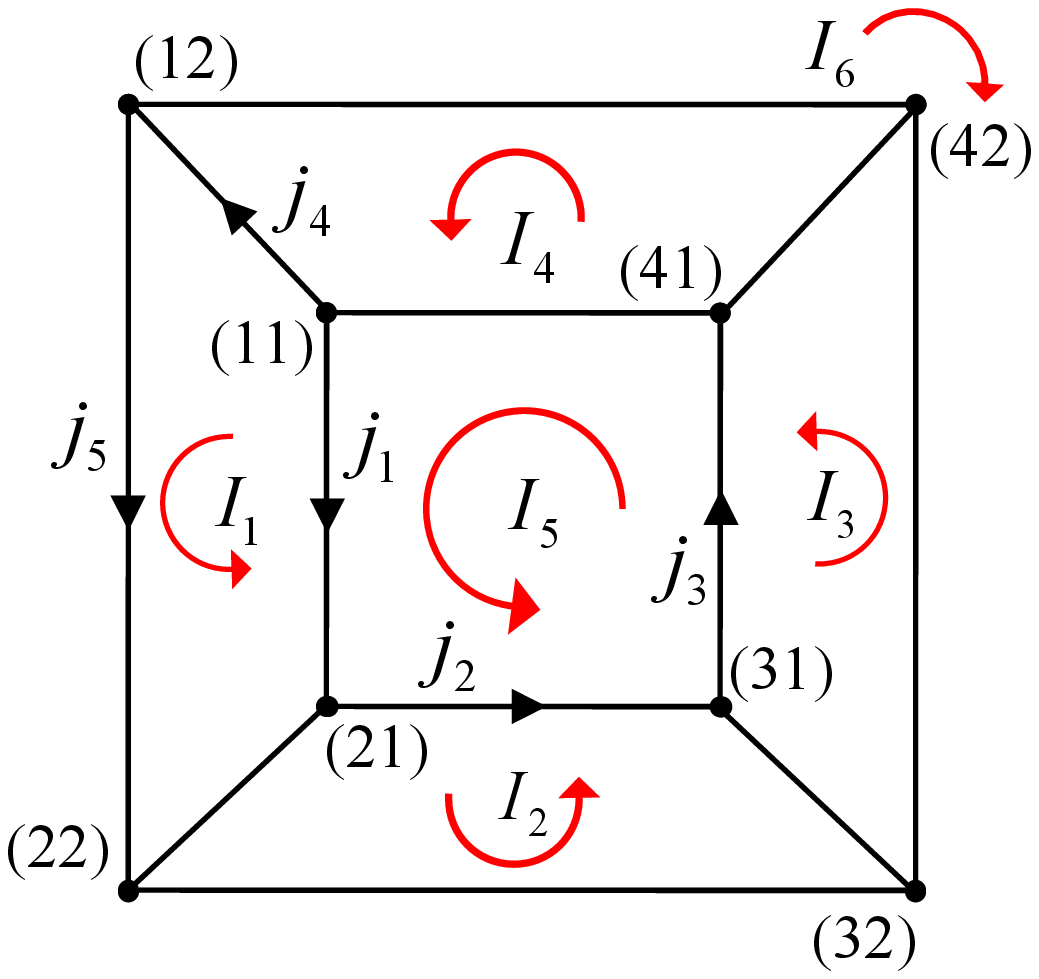}
\caption{Two dimensional projection of the 3D configuration space
of Fig. \ref{fig:schem1} showing the different probability current loops.}
\label{fig:schem2}
\end{figure}


Considering the case where the two systems are far from each other and
the deformations are small compared to the average length of the arms $L$, such that
$R \ll u \ll L \ll D$, we can set up a perturbative scheme to investigate the
effect of the hydrodynamic interactions \cite{3SS-perturb}. Solving the above linear system of ten
equations, we can find all the velocities and the forces, from which we can calculate
the average swimming velocity of the three-bead system
$V^{L}=\frac{1}{3} \left\langle v_a+v_b+v_c \right\rangle$ and that of the two-bead system
$V^{R}=\frac{1}{2} \left\langle v_d+v_e \right\rangle$.
To the leading order in perturbation theory, we find
\begin{eqnarray}
V^{L}=\frac{7}{12}\frac{R}{L^2}\left\langle
\dot{u}_{1}^{L}u_{2}^{L}\right\rangle-\frac{1}{2}\frac{R L}{D^3} \left[\left\langle
u_{1}^{L} \dot{u}^{R}\right\rangle-\left\langle u_{2}^{L} \dot{u}^{R}\right\rangle\right],&&
\label{eq:VL}\\
V^{R}=\frac{R L}{D^3}\left[-2\left\langle\dot{u}_{1}^{L}u_{2}^{L}\right\rangle
+\frac{3}{2}\left\langle u_{1}^{L} \dot{u}^{R}\right\rangle+\frac{3}{2}\left\langle u_{2}^{L} \dot{u}^{R}\right\rangle\right].&&
\label{eq:VR}
\end{eqnarray}
We can also extract the average forces acting on the beads.
To the leading order in deformations, this yields:
\begin{eqnarray}
\langle f_{a} \rangle=\frac{5}{4}\pi\eta\frac{R^2}{L^2}\left\langle\dot{u}_{1}^{L}u_{2}^{L}\right\rangle
-3\pi\eta\frac{R^2 L}{D^3} \left[\left\langle u_{1}^{L} \dot{u}^{R} \right\rangle
+2 \left\langle u_{2}^{L} \dot{u}^{R}\right\rangle \right],\nonumber\\
\langle f_{c} \rangle=\frac{5}{4}\pi\eta\frac{R^2}{L^2}\left\langle\dot{u}_{1}^{L}u_{2}^{L}\right\rangle
+3\pi\eta\frac{R^2 L}{D^3} \left[2 \left\langle u_{1}^{L} \dot{u}^{R} \right\rangle
+\left\langle u_{2}^{L} \dot{u}^{R}\right\rangle \right],\nonumber\\
\langle f_{d} \rangle=9\pi\eta\frac{R^2 L}{D^3} \left[\left\langle u_{1}^{L} \dot{u}^{R} \right\rangle
+\left\langle u_{2}^{L} \dot{u}^{R}\right\rangle \right].\hskip 3.1cm\nonumber
\end{eqnarray}
Note that $\langle f_{b} \rangle=-\langle f_{a} \rangle-\langle f_{c} \rangle$
and $\langle f_{e} \rangle=-\langle f_{d} \rangle$.

The above expressions for the velocities and forces are given in terms of the three average quantities $\left\langle\dot{u}_{1}^{L}u_{2}^{L}\right\rangle$, $\left\langle u_{1}^{L} \dot{u}^{R} \right\rangle$, and $\left\langle u_{2}^{L} \dot{u}^{R}\right\rangle$, which correspond to the average rates of sweeping
enclosed areas in the three perpendicular sections of the three dimensional ($u_{1}^{L}$, $u_{2}^{L}$, $u^{R}$)
configuration space of the system, respectively. For {\em deterministic} conformational changes of the form
$$u_{1}^{L}=d \cos (\Omega t-\varphi_{1}^{L}),$$
$$u_{2}^{L}=d \cos (\Omega t-\varphi_{2}^{L}),$$
$$u^{R}=d \cos (\Omega t-\varphi^{R}),$$
we can calculate them using time averaging over a period.
This yields
\begin{eqnarray}
\left\langle\dot{u}_{1}^{L}u_{2}^{L}\right\rangle=\frac{1}{2} d^2 \Omega \sin(\varphi_{1}^{L}-\varphi_{2}^{L}),
\label{eq:phi12} \\
\left\langle u_{1}^{L} \dot{u}^{R} \right\rangle=\frac{1}{2} d^2 \Omega \sin(\varphi^{R}-\varphi_{1}^{L}),
\label{eq:phiR1} \\
\left\langle u_{2}^{L} \dot{u}^{R}\right\rangle=\frac{1}{2} d^2 \Omega \sin(\varphi^{R}-\varphi_{2}^{L}).
\label{eq:phiR2}
\end{eqnarray}
The above equations manifestly show that the relative importance of these three conformational space area-sweeping rates is determined by the relative phases of the deformations. We now aim to address the question of whether such a concept can exist at small scales where the conformational changes are stochastic.

\section{Stochastic Systems}
To construct a {\em statistical} theory for the deformations 
of the two systems we assume that they have distinct conformational states and the
deformations can be modeled as stochastic jumps between these states that occur at given
rates \cite{RG+AA,RG+AA-2}. The three-sphere system can be described with four
states and the two-sphere system with two states, which make a total of eight
distinct conformational states in the three dimensional configuration space,
as shown in Fig. \ref{fig:schem1}. More specifically the states of the three-sphere swimmer
are labeled by the index $i$ as follows: $i=1$ the two arms are closed ($u_{1}^{L}=0, u_{2}^{L}=0$),
$i=2$ the right arm is open ($u_{1}^{L}=\delta, u_{2}^{L}=0$), $i=3$ the two arms are open
($u_{1}^{L}=\delta, u_{2}^{L}=\delta$), and $i=4$ the left arm is open
($u_{1}^{L}=0, u_{2}^{L}=\delta$). For the two-sphere system, we only have
two possibilities: $\alpha=1$ the arm is closed ($u^{R}=0$) and
$\alpha=2$ it is open ($u^{R}=\delta$). To describe the instantaneous
state of the system we denote the probability of finding the left swimmer at
state $i$ ($i=1,\cdots,4$) and the right two-bead system at state $\alpha$
($\alpha=1,2$) by $P_{i\alpha}$. These probabilities are normalized as
$$\sum_{i,\alpha}P_{i\alpha}=1.$$
The kinetics of the conformational transitions of the
two coupled systems is given by introducing the corresponding transition rates.

We assume that the conformational changes happen one at a time, which means
that transitions are only allowed between states that are nearest neighbors in the
cubic configuration space shown in Fig. \ref{fig:schem1}. We denote the transition
rate for the jump from state $i$ to state $j$ for the left swimmer when the
two-bead system is in state $\alpha$ by $k_{ji}^{L}(\alpha)$. Similarly, the
transition rate for the two-bead system jumping from state $\alpha$ to state $\beta$
when the three-sphere system is in state $i$ is denoted as $k_{\beta\alpha}^{R}(i)$.
Note that the rates for conformational changes within each system in principle depend
on the state of the other system. The cubic configuration space has six current loops
corresponding to six faces, as shown in Fig. \ref{fig:schem2}. These currents, however,
are subject to an overall conservation law, which implies that only five independent
currents exist in the system. We define the following currents
\begin{eqnarray}
j_1&=&P_{11}k_{21}^{L}(1)-P_{21}k_{12}^{L}(1),\nonumber\\
j_2&=&P_{21}k_{32}^{L}(1)-P_{31}k_{23}^{L}(1),\nonumber\\
j_3&=&P_{31}k_{43}^{L}(1)-P_{41}k_{34}^{L}(1),\nonumber\\
j_4&=&P_{11}k_{21}^{R}(1)-P_{12}k_{12}^{R}(1),\nonumber\\
j_5&=&P_{12}k_{21}^{L}(2)-P_{22}k_{12}^{L}(2),\nonumber
\end{eqnarray}
in terms of the probabilities and the rates, and can use them
to calculate the currents in the loops as follows (see Fig. \ref{fig:schem2}):
\begin{eqnarray}
I_1&=&\frac{1}{6} (-3 j_1 + j_2 + j_3 + j_4 + j_5),\nonumber\\
I_2&=&\frac{1}{6} (3 j_1 - 5 j_2 + j_3 + j_4 + j_5),\nonumber\\
I_3&=&\frac{1}{6} (3 j_1 + j_2 - 5 j_3 + j_4 + j_5),\nonumber\\
I_4&=&\frac{1}{6} (-3 j_1 + j_2 + j_3 - 5 j_4 + j_5),\nonumber\\
I_5&=&\frac{1}{6} (3 j_1 + j_2 + j_3 + j_4 + j_5),\nonumber\\
I_6&=&-I_1-I_2-I_3-I_4-I_5.\nonumber
\end{eqnarray}
We can now solve the steady state master equation for the system and
calculate the probabilities and the currents.

\begin{figure}[t]
\includegraphics[width=.90\columnwidth]{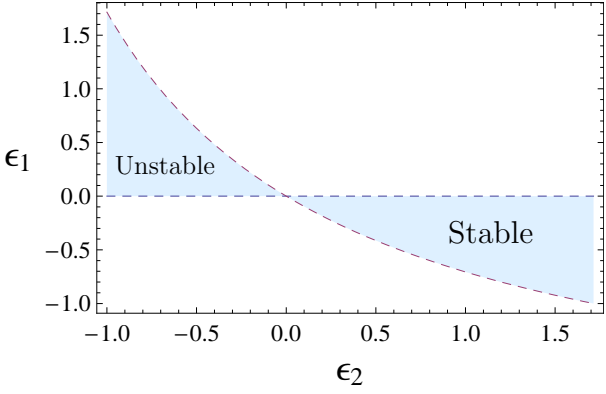}
\caption{Phase diagram of possible states.
Steady state solutions in the form of bound states with constant
separation ($\dot{D}=0$) are highlighted, and the Stable and Unstable
regions are shown.}
\label{fig:plot}
\end{figure}

Using the currents in the loops, we can write down expressions for the average
rates of sweeping areas in the three perpendicular sections of the
configuration space, as shown in Fig. \ref{fig:schem2}. The results, which are
the statistical analogs of eqs. (\ref{eq:phi12}), (\ref{eq:phiR1}), and (\ref{eq:phiR2}),
read
\begin{eqnarray}
\left\langle\dot{u}_{1}^{L}u_{2}^{L}\right\rangle=\delta^2 (I_6-I_5),
\label{eq:I56} \\
\left\langle u_{1}^{L} \dot{u}^{R} \right\rangle=\delta^2 (I_3-I_1),
\label{eq:I31} \\
\left\langle u_{2}^{L} \dot{u}^{R}\right\rangle=\delta^2 (I_4-I_2).
\label{eq:I42}
\end{eqnarray}
Due to the sign convention used in the definition of the currents the current running
through opposite faces of the cube in Fig. \ref{fig:schem2} have opposite signs.
Therefore, the total rate of sweeping a certain projected area in the configuration
space is the difference between the currents of the corresponding opposite faces in
the cube, as eqs. (\ref{eq:I56}), (\ref{eq:I31}), and (\ref{eq:I42}) show. The
area of each projection is equal to $\delta^2$. The above expressions can be used in
eqs. (\ref{eq:VL}) and (\ref{eq:VR}) to calculate the average swimming velocities of
the two systems.

In the most general case with arbitrary rates the explicit form of the
resulting probabilities and velocities is cumbersome and therefore not
shown here. To illustrate the generic features of the solution, we focus on a
simplified example with the following choices for the transition rates. For transitions
in the three-sphere system we choose
\begin{equation}
k_{ij}^{L}(\alpha)=\left\{
\begin{array}{l}
(1+\epsilon_1)\omega,~~ i=1,j=2,\alpha=1, \nonumber\\
\\
(1+\epsilon_2)\omega,~~ i=1,j=2,\alpha=2, \nonumber\\
\\
\omega,~~~~~~~~~~~ {\rm other~states}
\end{array}
\right.
                   \label{G-solution}
\end{equation}
For the two-sphere system, we choose
$$k_{\beta\alpha}^{R}(i)=\omega,~~~{\rm for~all~states.}$$
The above choices allow us to only focus on the effect of the correlation between the two devices, as
having different values for $\epsilon_1$ and $\epsilon_2$ means that the rate of the three-sphere system going from state 1 to state 2, which means opening its right arm, depends on whether the two-sphere system is in the closed or the open state.
If $\epsilon_1=\epsilon_2=0$, detailed balance holds and neither of the two components
has a net motion. With the above choices, the average velocities of the two systems
can be found as
\begin{eqnarray}
&&V^{L}=R \omega\left(\frac{\delta}{L}\right)^2 \left[\frac{7}{24} y_0-\frac{1}{2}\left(\frac{L}{D}\right)^3 y_1\right],\label{eq:VL2}\\
&&V^{R}=R \omega\left(\frac{\delta}{L}\right)^2 \left(\frac{L}{D}\right)^3 \left[-y_0+\frac{3}{2} y_1\right],\label{eq:VR2}
\end{eqnarray}
where
\begin{eqnarray}
y_0&=&\frac{12 (\epsilon_1+\epsilon_2)+5 \epsilon_1\epsilon_2}
{56 (\epsilon_1+\epsilon_2)+15 \epsilon_1\epsilon_2+192},\label{eq:x0}\\
y_1&=&\frac{6 (\epsilon_2-\epsilon_1)}
{56 (\epsilon_1+\epsilon_2)+15 \epsilon_1\epsilon_2+192}.\label{eq:x1}
\end{eqnarray}
The term proportional to $y_0$ in eq. (\ref{eq:VL2}) is the spontaneous swimming velocity
of the three-sphere system, while the $y_0$ contribution in eq. (\ref{eq:VR2}) is the passive
velocity at the location of the two-sphere system caused by the swimming of the three-sphere system.
The contributions proportional to $y_1$ in eqs. (\ref{eq:VL2}) and (\ref{eq:VR2}) are active contributions
originating from a coherence between $u_{1}^{L}$ and ${u}^{R}$ in the form of $\left\langle u_{1}^{L} \dot{u}^{R} \right\rangle \neq 0$. Note that the coherence will disappear when $\epsilon_1=\epsilon_2$.

An interesting consequence of the coherent coupling is that the two systems can form a moving hydrodynamic bound state at a fixed separation. Noting that $\frac{{\rm d}}{{\rm d} t} D(t)=V^{R}-V^{L}$ or
\begin{equation}
\frac{{\rm d} D(t)}{{\rm d} t}=R \omega\left(\frac{\delta}{L}\right)^2 \left[\left(\frac{L}{ D(t)}\right)^3 \left(2 y_1-y_0\right)-\frac{7}{24} y_0\right],
\end{equation}
we can find the conditions at which stable and unstable bound states are possible in the effective dynamical equation for $D(t)$, as shown in Fig. \ref{fig:plot}. The equilibrium distance between the two systems in the stable hydrodynamic bound states is given as
\begin{equation}
D_{\rm eq}=L \left(\frac{24}{7}\right)^{1/3}\left[\frac{-\epsilon_1 (24+5 \epsilon_2)}{12 (\epsilon_1+\epsilon_2)+5 \epsilon_1\epsilon_2}\right]^{1/3},\label{eq:Deq}
\end{equation}
which can be controlled by changing the transition rates. We note that the hydrodynamic-induced formation of bound states of a pair of microorganisms has been recently observed experimentally \cite{RaymondPRL}.

\section{Discussion}
Our analysis shows that the concept of relative internal phase and coherence between a number of systems that undergo stochastic deformations in a hydrodynamic medium at low Reynolds number is well defined. Stochastic coherence could result from average correlations that can be induced between various modes of the conformational transitions, and need not exist instantaneously to lead to average correlated behavior. Comparison between eqs. (\ref{eq:phi12}), (\ref{eq:phiR1}), and (\ref{eq:phiR2}) that are defined for deterministic systems and eqs. (\ref{eq:I56}), (\ref{eq:I31}), and (\ref{eq:I42}) that are defined for stochastic systems shows how average relative phase between various modes of the conformational transitions can be defined and probed in terms of the currents in the configuration space of the system.

In the present study, coherence between the two subsystems is introduced via the rates defined in eq. (\ref{G-solution}). When $\epsilon_1 \neq \epsilon_2$, the rate of opening of the right arm of the three-sphere swimmer is chosen to depend on whether the two-sphere system is in the closed or open state; it is exactly this difference that leads to the correlation term proportional to $y_1$ in eqs. (\ref{eq:VL2}) and (\ref{eq:VR2}), as can be seen from the explicit dependence of $y_1 \propto (\epsilon_2-\epsilon_1)$ in eq. (\ref{eq:x1}). This means that while it is possible to have coherence between different parts of the system when undergoing stochastic fluctuations, this coherence still needs to be imposed via the different rates in the kinetic equations. Physically, what this means is that the coherence needs to introduced in the system via correlations between different conformational states of the system and the rates of transitions between them. The correlations are reminiscent of the allosteric interactions between proteins \cite{stryer}, and could in principle be engineered for artificial systems using similar strategies.
These strategies could involve physical interactions arising from electrostatic forces {\em etc}, hydrodynamic interactions, and other effects that could modify the transition rates by introducing additional mechanical energy contributions (costs) in the deformation process and hence affecting the transition rates via the exponential (Arrhenius) dependence on energy change. Alternatively, these correlations could be induced via external means such as laser pulses that would affect transition rates only in certain conformational states. However they are enforced, the present study asserts that such correlations could lead to a sustainable notion of coherence between
stochastically fluctuating nano-scale devices in water.

We have considered the transition rates for the conformational changes of the two small systems to be
time independent. If this assumption is not valid for any reason, the time dependence in the rates could
weaken the degree of coherence in the system and ultimately fully eliminate it if it is sufficiently strong. This is equivalent to introducing time dependence in the phases in the deterministic case [described by eqns. (\ref{eq:phi12}), (\ref{eq:phiR1}), and (\ref{eq:phiR2})], which could destroy the phase coherence.
Such a time dependence could occur due to temperature fluctuations \cite{landau}, which in local thermodynamics approximation could affect the transition rate through a dependence of the form
\begin{equation}
\omega\sim {\rm exp}\left[-\frac{f \delta}{k_{\rm B}T}\right],\label{eq:rate}
\end{equation}
where $f$ represents a typical force involved in the conformational change.
We can estimate the magnitude of $f$ using the typical drag force experienced by a sphere
of radius $R$ and moving velocity $v=\delta \omega$, namely $f\approx 6\pi\eta R \delta^2 \omega$.
In the present work we have neglected temperature fluctuations. To examine the validity of this
assumption, we can use the complementarity relation $\Delta E~\Delta(1/T)\approx -k_{\rm B}$,
which relates the strength of energy and temperature fluctuations, to estimate the magnitude of
temperature fluctuations as $\Delta T^2=k_{\rm B} T^2/C$, where $C$ is the heat capacity \cite{landau}. Using this simplified picture, we can estimate the effect of temperature fluctuations on
the transition rates via
\begin{equation}
\frac{\Delta\omega}{\omega}=\frac{f\delta}{k_{\rm B}T} \times
\frac{\Delta T}{T}\approx\frac{6\pi\eta R \delta^2 \omega}{k_{\rm B}T} \sqrt{\frac{k_{\rm B}}{C}}.
\label{delta-1}
\end{equation}
In order to have $\Delta\omega/\omega \ll 1$, which would guarantee that the above assumption
is valid (i.e. temperature fluctuations can be ignored) the following condition must hold:
\begin{equation}
\omega \ll \frac{k_{\rm B}T}{6\pi\eta R \delta^2} \sqrt{\frac{C}{k_{\rm B}}}.
\label{delta-1}
\end{equation}
Putting $R=\delta=1$ nm at room temperature for water, we find the following condition:
\begin{equation}
\omega \ll  \left(10^{8} \;{\rm s}^{-1}\right) \times \sqrt{\frac{C}{k_{\rm B}}}.
\label{delta-2}
\end{equation}
While there is a debate in the literature about the correct choice for $C$, namely whether
it is the heat capacity of the entire system \cite{mandelbort} or that of the subsystem only \cite{TFlucExper}, eqns. (\ref{delta-1}) and (\ref{delta-2}) show that in our case neglecting
temperature fluctuations is justified either way. Note also that our analysis has ignored quantum
fluctuations, which means that the following criterion should also hold:
$\omega \ll \frac{k_{\rm B}T}{\hbar}$.

Finally, we note that we have made the specific choice of a three-sphere system and a two-sphere system, because its configuration space is 3D and can be easily visualized (Figs. \ref{fig:schem1} and \ref{fig:schem2}). The same analysis can be easily generalized to the case of two three-sphere swimmers, in which case
the configuration space will be 4D and the bookkeeping of the projected areas in the graph where the probability currents are flowing is slightly more delicate.

In conclusion, we have shown that stochastic swimmers could actively couple to each other using hydrodynamic interactions. As an example, we demonstrated that the coupling could be tailored such that an active swimmer could hunt a non-swimming system into a stable moving bound state, by ``tuning'' into the right correlated transition rates. Our results show that the notion of internal phase and coherence could be important even for fluctuating systems that are coupled via hydrodynamic interactions at the nano-scale.

\acknowledgments
We acknowledge financial support from MPIPKS (A.N.), CNRS (R.G.), and EPSRC (A.N. and R.G.).


\begin{thebibliography}{0}

\bibitem{PoLa-rev}
Lauga E. and Powers T. R., Rep. Prog. Phys. {\bf 72} (2009) 096601.

\bibitem{Taylor}
Taylor G. I., Proc. R. Soc. A {\bf 209} (1951) 447.

\bibitem{Purcell}
Purcell E. M., Am. J. Phys. {\bf 45} (1977) 3.

\bibitem{shapere}
Shapere A. and F. Wilczek, Phys. Rev. Lett. {\bf 58} (1987) 2051.

\bibitem{becker}
Becker L. E., Kohler S. A., Stone H. A., J. Fluid Mech. {\bf 490} (2003) 15.

\bibitem{3SS}
Najafi A. and Golestanian R., Phys. Rev. E {\bf 69} (2004) 062901.

\bibitem{josi}
Avron J. E.,  Gat O. and Kenneth O., Phys. Rev. Lett. {\bf 93} (2004) 186001.

\bibitem{drey}
Dreyfus R., Baudry J. and Stone H. A., Eur. Phys. J. B {\bf 47} (2005) 161.

\bibitem{feld}
Felderhof B. U., Phys. Fluids {\bf 18} (2006) 063101.

\bibitem{peko}
Tam D. and Hosoi A. E., Phys. Rev. Lett. {\bf 98} (2007) 068105.

\bibitem{yeomans1}
Earl D. J., Pooley C. M., Ryder J. F., Bredberg I., and Yeomans J. M.,
J. Chem. Phys. {\bf 126} (2007) 064703.

\bibitem{Yeomans2}
Pooley C. M., Alexander G. B., and Yeomans J., Phys.
Rev. Lett. {\bf 99} (2007) 228103.

\bibitem{Karsten}
G\"unther S. and Kruse K., Europhys. Lett. {\bf 84} (2008) 68002.

\bibitem{3SS-perturb}
Golestanian R. and Ajdari A., Phys. Rev. E {\bf 77} (2008) 036308;
Golestanian R., Euro. Phys. J. E {\bf 25} (2008) 1 (R).

\bibitem{zargar}
Zargar R., Najafi A. and Miri M. F., Phys. Rev. E {\bf 80} (2009) 026308.

\bibitem{BergEC}
Berg H. C., {\em E. coli in Motion} (Springer-Verlag, New York)
2004.

\bibitem{Kruse}
Riedel I. H. {\em et al.}, Science {\bf 309} (2005) 300.

\bibitem{bray}
Bray D., {\em Cell Movements: From Molecules to Motility} 2nd ed.
(Garland, New York) 2001.

\bibitem{exp-swim}
Dreyfus R. {\em et al.}, Nature {\bf 437} (2005) 862;
Tierno P. {\em et al.}, Phys. Rev. Lett. {\bf 101} (2008) 218304;
J. Phys. Chem. B {\bf 112} 16525 (2008);
Leoni M. {\em et al.}, Soft Matter {\bf 5} (2009) 472.

\bibitem{Darnton}
Darnton N., Turner L., Breuer K., and Berg H. C., Biophys. J. {\bf 86} (2004) 1863.

\bibitem{gla}
Golestanian R., Liverpool T. B., and Ajdari A., Phys. Rev. Lett. {\bf
94} (2005) 220801; New J. Phys. {\bf 9} 126 (2007).

\bibitem{jon}
Howse J. R. {\em et al.}, Phys. Rev. Lett. {\bf 99} (2007) 048102.

\bibitem{Vlad}
Lobaskin V., Lobaskin D., and Kulic I., Eur. J. Phys. Spec.
Top. {\bf 157} (2008) 149.

\bibitem{Dunkel}
Dunkel J. and Zaid I. M., Phys. Rev. E {\bf 80} (2009) 021903.

\bibitem{PeMo}
Peruani F. and Morelli L. G., Phys. Rev. Lett. {\bf 99} (2007) 010602.

\bibitem{msd}
Golestanian R., Phys. Rev. Lett. {\bf 102} (2009) 188305.

\bibitem{RG+AA}
Golestanian R. and Ajdari A., Phys. Rev. Lett. {\bf 100} (2008) 038101.

\bibitem{ped-rev}
Pedley T. J. and Kessler J. O., Annu. Rev. Fluid Mech.
{\bf 24} (1992) 313.

\bibitem{simulation}
Hernandez-Oritz J. P.{\em et al.}, Phys. Rev. Lett. {\bf 95} (2005) 204501;
Ishikawa T. and T.J. Pedley, J. Fluid Mech. {\bf 588} (2007) 437;
Aranson I. S. {\em et al.}, Phys. Rev. E {\bf 75} (2007) 040901;
Saintillan D. and Shelley M. J., Phys. Rev. Lett. {\bf 100}  (2008) 178103.

\bibitem{instability}
Simha R. A. and Ramaswamy S., Phys. Rev. Lett. {\bf 89} (2002) 058101;
Hatwalne Y. {\em et al.}, Phys. Rev. Lett. {\bf 92} (2004) 118101;
Ahmadi A. {\em et al.}, Phys Rev E {\bf 74} (2006) 061913;
Baskaran A. and Marchetti M. C., Proc. Natl. Acad. Sci. (USA) {\bf 106} (2009) 15567.

\bibitem{Yeomans3}
Alexander G. P. and Yeomans J. M., Europhys. Lett. {\bf 83} (2008) 34006.

\bibitem{denis}
Lauga E. and Bartolo D., Phys. Rev. E {\bf 78} (2008) 030901.

\bibitem{sync}
Lagomarsino M. C. {\em et al.}, Eur. Phys. J. B {\bf 26} (2002) 81�88;
Lagomarsino M. C. {\em et al.}, Phys. Rev. E {\bf 68} (2003) 021908;
Kim M. {\em et al.}, Proc. Natl. Acad. Sci. (USA) {\bf 100} (2003) 15481;
Kim M. and T.R. Powers, Phys. Rev. E {\bf 69} (2004) 061910;
Reichert M., Stark H., Eur. Phys. J. E {\bf 17} (2005) 493;
Vilfan A. and J\"ulicher F., Phys. Rev. Lett. {\bf 96} (2006) 058102;
Kim Y. W. and Netz R., Phys. Rev. Lett. {\bf 96} (2006) 158101;
Guirao B. and Joanny J-F., Biophys. J. {\bf 92} (2007) 1900-1917;
Elgeti J., PhD thesis Universit\"at zu K\"oln (2007);
Goldstein R. E. {\em et al.}, Phys. Rev. Lett. {\bf 103} (2009) 168103;
Putz V. B. and J.M. Yeomans, J. Stat. Phys. {\bf 137} (2009) 1001;
Qian B. {\em et al.}, Phys. Rev. E {\bf 80} (2009) 061919;
Uchida N. and Golestanian R., Phys. Rev. Lett. {\bf 104} (2010) 178103;
Uchida N. and Golestanian R., Europhys. Lett. {\bf 89} (2010) 50011.

\bibitem{RG+AA-2}
Golestanian R. and Ajdari A., J. Phys.: Condens. Matter {\bf 21} (2009) 204104.

\bibitem{RaymondPRL}
Drescher K. {\em et al.}, Phys. Rev. Lett. {\bf 102} (2009) 168101.

\bibitem{stryer}
Berg J. M., Tymoczko J. L., and Stryer L., {\em Biochemistry}
(W.H. Freeman and Co., New York) 2002.

\bibitem{landau}
Landau L. D., and Lifshitz E. M., {\em Statistical Physics} (Pergamon, London)
1980.

\bibitem{mandelbort}
Mandelbort B. B., Phys. Today, {\bf 42} No.1 (1989) 71.

\bibitem{TFlucExper}
Chui T. C. P. {\em et al.}, Phys. Rev. Lett. {\bf 69} (1992) 3005.


\end{thebibliography}
\end{document}